# Pseudo-Linear Time-Invariant Magnetless Circulators Based on Differential Spatiotemporal Modulation of Resonant Junctions

Ahmed Kord, *Graduate Student Member, IEEE*, Dimitrios L. Sounas, *Senior Member, IEEE*, and Andrea Alù, *Fellow, IEEE*

*Abstract*— In this paper, we present voltage- and current-mode differential magnetless non-reciprocal devices obtained by pairing two single-ended (SE) circulators, each consisting of three first-order bandpass or bandstop *LC* filters, connected in either a wye or a delta topology. The resonant poles of each SE circulator are modulated in time with 120 deg phase-shifted periodic signals, resulting in synthetic angular-momentum biasing achieved through spatiotemporal modulation (STM). We tailor the two SE circulators to exhibit a constant 180 deg phase difference between their STM biases. Unlike conventional differential time-variant circuits, for which only the even or odd spurs are rejected, we show that the proposed configuration cancels out all intermodulation (IM) products, thus making them operate alike linear time-invariant (LTI) circuits for an external observer. In turn, this property enhances all metrics of the resulting circulator, overcoming the limitations of SE architectures, and improving insertion loss, impedance matching, bandwidth and noise figure. We show that this differential architecture also significantly relaxes the required modulation parameters, both in frequency and amplitude. We develop a rigorous small-signal model to guide the design of the proposed circuits and to get insights into their pseudo-LTI characteristics. Then, we validate the theory with simulations and measurements showing remarkable performance compared to the current state of the art of magnetless non-reciprocal devices.

## I. INTRODUCTION

CIRCULATORS are three-port non-reciprocal components, crucial to enable full-duplex communication [1]-[6], since they allow unidirectional signal transmission from the transmitter (TX) to the antenna (ANT), while isolating the receiver (RX) from self-interference. Commercial circulators are based on magnetic biasing of rare-earth ferrite materials [7]-[10], which are bulky, expensive and incompatible with standard integrated-circuit (IC) technologies. In order to remove the magnet, active circulators that utilize the intrinsic non-reciprocal properties of transisters were explored, yet these devices have never become popular since they suffer from fundamentally poor noise figure and power handling performance [11]-[14]. Recently, linear periodically time-varying (LPTV) circuits have been presented as an alternative approach towards magnetless non-reciprocity, which can achieve low loss, small noise figure, watt-level power handling, and other benefits [15]-[34]. In particular, [29]-[34] proposed a synthetic spatiotemporal modulation (STM) angular-momentum biasing of resonant junctions, which results in symmetric circulators with decent performance in many metrics. The main challenge in this approach is that it suffers from strong IM products in close proximity to the desired band, due to mixing between the RF input and the relatively low-frequency modulation signals. These products not only pose an interference problem to neighboring channels, but they also require large modulation parameters to achieve good performance – e.g., the modulation amplitude in [33] was larger than 10 Vpp – which prohibits their integration using submicron CMOS technologies. Furthermore, they effectively reduce the circulator's overall power handling in an actual full-duplex system, since they could saturate the RX front-end and drive the TX power amplifier into instability because of load-pull effects. More importantly, these products enforce a bound on the minimum insertion loss of about 3 dB [31]-[33], which weakens the argument of circulators compared to other reciprocal interfaces based on couplers or balanced duplexers [35], [36]. Therefore, the rejection of these products is pivotal, and as important as cancelling the fundamental harmonic of the TX signal at the RX port, to enable the use of STM circulators in commercial systems. Although filtering may sound a reasonable option, it is, in fact, far from being practical, since it suffers from many problems. Specifically, adding filters at the circulator's three ports increases the overall size and degrades the total insertion loss. It also imposes a restriction on the minimum modulation frequency, in order to relax the requirements on the sharpness of these filters, which, in turn, requires large modulation amplitude to maintain sufficient isolation [33]. This not only increases power consumption and further complicates integration, but it may be even impossible

Manuscript received on August 31, 2017.
The authors are with the Department of Electrical and Computer Engineering, University of Texas at Austin, Austin, TX 78712, U.S.A. A.A. is also with the Advanced Science Research Center, City University of New York, New York, NY 10031, U.S.A. (corresponding author: A. A., +1.512.471.5922; fax: +1.512.471.6598; e-mail: aalu@gc.cuny.edu). This work was supported by the Qualcomm Innovation Fellowship, the Air Force Office of Scientific Research, the Defense Advanced Research Projects Agency, Lockheed Martin, Silicon Audio, the Simons Foundation, and the National Science Foundation. A.A. is currently the Chief Technology Officer of Silicon Audio RF Circulator. The terms of this arrangement have been reviewed and approved by The University of Texas at Austin in accordance with its policy on objectivity in research.

to achieve with practical varactors. Moreover, RF filters are typically non-reconfigurable, hence STM circulators' tunability, which is an important feature in modern communication systems, is sacrificed.

Ref. [34] proposed a partial solution to these problems based on combining two SE circulators with anti-phase STM bias through RF baluns. Yet, the results in [34] were based on heuristic investigations without deep understanding of the ultimate capabilities of this architecture, nor a detailed simulation and experimental validation. In this paper, we address this issue by developing a rigorous theory for differential STM circulators and showing that they have far more interesting characteristics than the ones reported in [34]. In particular, we show that the suitable combination of two SE circulators in a differential configuration surprisingly results in the total *cancellation of all IM products* for excitation at any port and at any frequency, thus making these circulators essentially pseudo-LTI circuits. This property, in turn, alleviates the trade-off between the IM products and the modulation parameters (recall that in SE circulators, a minimum modulation frequency is required to keep the IM products below a certain level [32], [33]), enabling a strong reduction of both modulation frequency and amplitude, while still achieving remarkable performance. Interestingly, this implies that the differential STM bias synthesizes a truly *continuous* angular momentum, mimicking mechanical motion that does not use temporal modulation [28], even though it is implemented using a discrete number of resonators (three per junction). This is fully consistent with magnetic-biased circulators, in which the aligned electron dipole moments imitate a *continuous* rotation of the ferrite disk at a macroscopic level, while they are in fact a collection of quantized spins [7]. Furthermore, in analogy with passive mixers [37], we introduce two dual implementations of differential STM circulators: (i) voltage-mode and (ii) current-mode. In voltage-mode circulators, which are implemented using bandstop/delta junctions, baluns are used to subtract the IM products, since they arise as *common voltages* at the terminals of the constituent SE circulators [33], [34]. Conversely, in current-mode circulators, which are based on bandpass/wye junctions, IM products arise as *differential currents*, and they can be rejected by tying the terminals of two SE circulators together to sum up and cancel these currents.

This paper is organized as follows. In Section II, we briefly summarize the SE STM circulators presented in [32], [33] and discuss their limitations. In Section III, we qualitatively analyze the proposed voltage- and current-mode topologies to explain how these circuits have emerged and to understand their differences. Then, we develop a detailed small-signal model for the voltage-mode architecture, prove that it is indeed IM-free, and derive analytical expressions for the S-parameters. For completeness, similar analysis for the current-mode topology is provided in the appendices. In Section IV, we present simulated and measured results for the voltage-mode topology with remarkable performance nearly in all metrics compared to the current state of the art. Finally, we draw our conclusions in Section V.

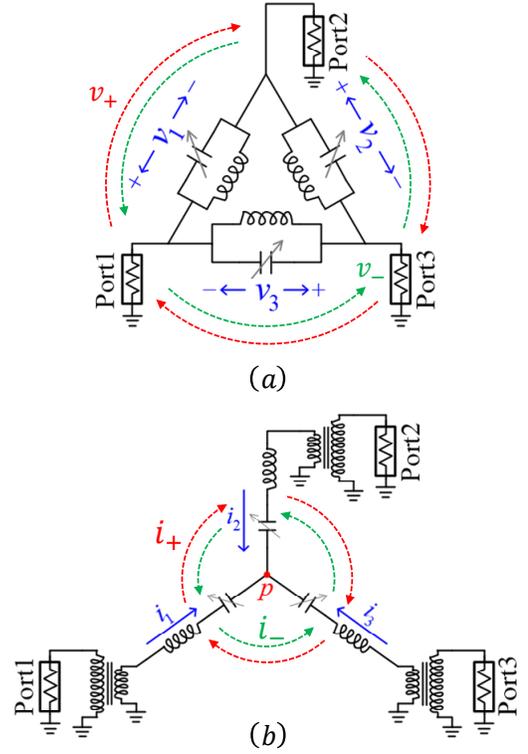

Fig. 1. Single-ended STM circulators: (a) Bandstop/delta topology. (b) Bandpass/wye topology.

## II. LIMITATIONS OF SINGLE-ENDED STM CIRCULATORS

The underlying physical principle of STM circulators is the same as for magnetic circulators, i.e. the degeneracy between two counter-rotating modes in a resonant junction is lifted by the applied bias. At radio frequencies, this was achieved by modulating the natural oscillation frequencies of bandpass or bandstop $LC$ tanks connected in a wye or a delta topology, respectively, with 120 deg phase-shifted periodic signals having the same frequency and amplitude [32], [33]. Fig. 1 shows a simplified schematic for both topologies, where the resonance frequency of the $n$-th tank is given by

$$f_n = f_0 + kV_m \cos(\omega_m t + \varphi_n), \quad n = 1, 2, 3 \qquad (1)$$

where $n$ is the tank index, $f_0$ is the static unmodulated resonance frequency of all tanks, $V_m$ and $\omega_m = 2\pi f_m$ are the modulation amplitude and angular frequency, respectively, $k$ is a constant with Hz/Volt units quantifying the effect of the modulation voltage on the resonance frequencies of the tanks, and $\varphi_n = (n-1)\alpha$, where $\alpha = 120°$. Such modulation scheme results in what we call STM angular-momentum biasing, since it involves phase variation in space ($\varphi$ direction) and in time. The bias direction is, by definition, the direction along which the phase of the modulation signals increases. For example, (1) assumes that the phase increases in the clock-wise direction based on the port definitions of Fig. 1, thus resulting in a clock-wise STM bias. This, in turn, provides a preferred sense of precession for the counter-rotating modes of the resonant junctions in Fig. 1. These modes can be defined if we express the tank voltages $v_n$ in Fig. 1(a) or, similarly, the tank currents $i_n$ in Fig. 1(b) as a superposition of two quantities as follows

$$v_n = v_+ e^{j(n-1)\alpha} + v_- e^{-j(n-1)\alpha}, \quad (2)$$
$$i_n = i_+ e^{j(n-1)\alpha} + i_- e^{-j(n-1)\alpha}, \quad (3)$$

where $v_\pm e^{\pm j(n-1)\alpha}$ and $i_\pm e^{\pm j(n-1)\alpha}$ are the aforementioned counter-rotating modes. Notice that the phase of these modes increases either clockwise (+) or counter clockwise (−), and it adds up to 360 deg in one cycle. In general, a common mode $v_0$ and $i_0$ whose phase is the same for all tanks may also be defined, yet it was shown in [33] that the cancellation of this mode is necessary to have optimal performance for STM circulators. The circuits in Fig. 1 indeed satisfy this condition since a non-zero common mode in either topologies would violate Kirchhoff's laws. To avoid confusion with the common and differential components decomposition of the total signal at the RF ports, $v_0$ and $i_0$ will be called the *in-phase* mode in the rest of this paper.

The topologies in Fig. 1 achieve strong non-reciprocity without magnets, yet they suffer from several disadvantages. In particular, the wye topology requires many filters in the modulation network, which increases the overall form factor and complicates the design. It also requires using impedance transformers at the 50 Ohm ports to increase the loaded quality factor $Q_l$ of the resonant junction, which is necessary to achieve strong non-reciprocity [32]. Clearly, these transformers add more intrinsic loss and increase the size further. Moreover, the constituent series $LC$ tanks amplify the input RF voltage across the varactors roughly by the same order as $Q_l$, which degrades the circulator's linearity and power handling. These problems can be overcome with the delta topology, which however requires a large modulation amplitude (>10 Vpp in [33]) to achieve good performance, thus increasing the dynamic power consumption and prohibiting integration in submicron CMOS technologies. Another serious problem that both topologies suffer from is the strong IM products due to mixing between RF and modulation frequencies. As mentioned in the introduction, these products limit the performance, particularly power handling and insertion loss. In the next sections, we present a differential architecture that overcomes all these problems and results in remarkable performance compared to the current state of the art.

### III. Theory and Proposed Circuit Topology

#### A. Voltage- and Current-mode Differential Topologies

In order to understand the operation principle of differential STM circulators, consider adding a constant phase $\theta$ to all modulation signals in (1), i.e., $\varphi_n = (n-1)\alpha + \theta$. One can prove, following the analysis in [33], that the fundamental component of the rotating modes, and subsequently the S-parameters, remain exactly the same, while the IM products at $\omega \pm \omega_m$ become

$$V_\pm(\omega \mp \omega_m, \omega) = V_\pm(\omega \mp \omega_m, \omega)\big|_{\theta=0°} e^{\pm j\theta}, \quad (4)$$
$$I_\pm(\omega \mp \omega_m, \omega) = I_\pm(\omega \mp \omega_m, \omega)\big|_{\theta=0°} e^{\pm j\theta}, \quad (5)$$

where $V_\pm(\omega \mp \omega_m, \omega)$ and $I_\pm(\omega \mp \omega_m, \omega)$ are the Fourier transforms of the generated IM products due to an input excitation at $\omega$. The expressions of $V_\pm(\omega \mp \omega_m, \omega)\big|_{\theta=0°}$ and $I_\pm(\omega \mp \omega_m, \omega)\big|_{\theta=0°}$ as a function of the circuit elements and modulation parameters can be found in [32] and [33], respectively. This finding suggests that combining two SE circulators with a phase difference $\Delta\theta = 180°$ between the modulation signals of the constituent SE circulators cancels these products entirely.

Based on this observation, Fig. 2(a) and Fig. 2(b) show the proposed voltage- and current-mode differential architectures, respectively. The constituent SE circulators in the voltage-mode topology are based on bandstop/delta junctions (see Fig. 1(a)) and are combined together using differential ports (or baluns). In contrast, the current-mode topology employs bandpass/wye junctions (see Fig. 1(b)) with their terminals directly tied together. Both topologies ensure that the IM products of each SE circulator have opposite parities, hence they destructively interfere at all ports, while the fundamental components are in phase and, therefore, sum up constructively. More specifically, the voltage-mode topology yields even symmetry for the IM products, therefore resulting in an infinite effective port impedance for these products, as shown in Fig. 3(a). On the other hand, the fundamental components experience a virtual ground at the middle of the differential ports, leading to an effective port impedance that is half of the original differential port impedance $Z_0$, as shown in Fig. 3(b). Unlike the IM products, the finite port impedance at the fundamental frequency allows current flow and, consequently, power transfer to the ports. From duality, the current-mode topology results in opposite symmetries, i.e., odd for the IM and even for the fundamental frequencies, as shown in Fig. 3(c) and Fig. 3(d), respectively.

Fig. 3(a) and Fig. 3(c) show that the IM products in the voltage- or current-mode topologies are prohibited from leaking into the external ports since they *see* an effective open or short circuit impedance, respectively. Since power is conserved, these products reflect back to the resonant junctions and they can be regarded as new excitations at $\omega \pm \omega_m$, which mix with the modulation signal, similarly to the original input at $\omega$, and generate new harmonics at $\omega \pm \omega_m$, $\omega$, and $\omega \pm 2\omega_m$. The second-order spurs at $\omega \pm 2\omega_m$ exhibit the same symmetry as the fundamental component at $\omega$, hence their half-circuit models are the same as Fig. 3(b)-(d) and, therefore, can leak to the external ports. In general, the circulator is designed to be matched at the fundamental harmonic but this does not guarantee matching at $\omega \pm 2\omega_m$, hence the IM products at these frequencies partially reflect and enter the circulator again.

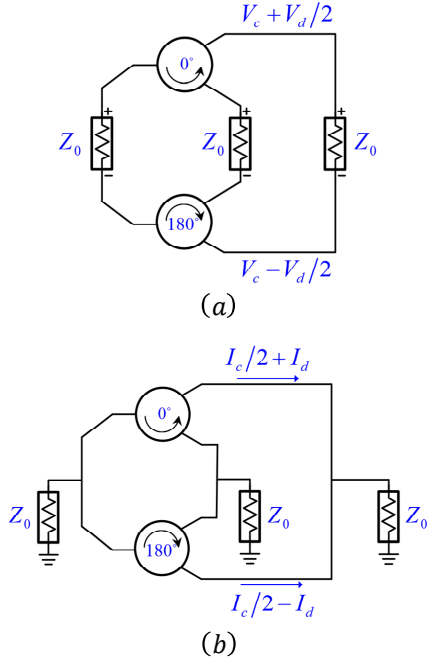

Fig. 2. Differential STM circulator: (a) Voltage-mode topology. (b) Current-mode topology.

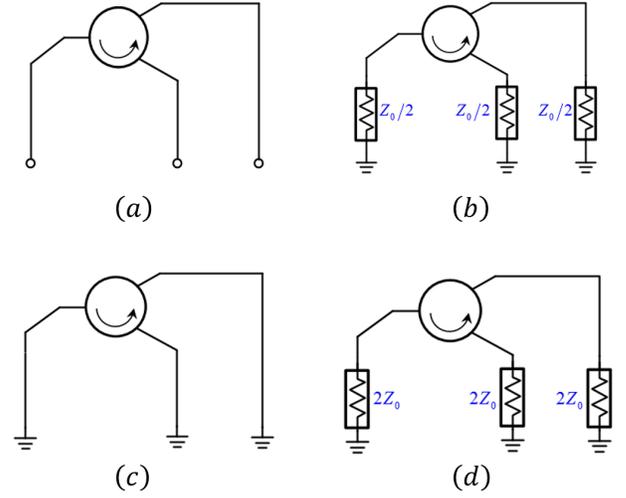

Fig. 3. Half-circuit models for: (a) Voltage-mode fundamental harmonic. (b) Voltage-mode IM products. (c) Current-mode fundamental harmonic. (d) Current-mode IM products.

Similarly, this gives rise to new harmonics at $\omega \pm 3\omega_m$, which have the same symmetry as the first-order products at $\omega \pm \omega_m$ and therefore are completely reflected back to the resonant junctions. By continuing this recursive process, we find that the spectrum at the RF ports is expected to contain the even-order harmonics $\omega$, $\omega \pm 2\omega_m$, $\omega \pm 4\omega_m$, …. This qualitative analysis suggests that the performance improvement due to the differential configuration would be incremental, and only associated with a weakening of the first non-zero higher products at $\omega \pm 2\omega_m$. In the next section, however, we prove that both proposed circulators in Fig. 2 do not generate any IM product, making them pseudo-LTI circuits and drastically improving the performance. This result cannot be predicted from the half-circuit models of Fig. 3, which are typically developed for conventional differential circuits, and it requires a rigorous analysis of the complete circuit. Since the analysis of both topologies is quite similar, we focus in the rest of this paper on the voltage-mode architecture, we develop its small-signal model (detailed analysis in Appendix A), and provide simulated and measured results with remarkable performance. For completeness, analysis of the current-mode topology is provided in Appendix B.

### B. Linear Small-Signal Analysis of Voltage-mode Topology

In this section, we provide a theoretical analysis for the voltage-mode topology. Fig. 4(a) shows the complete circuit implementation of the voltage-mode STM circulator where $LC$ baluns ($L_{rf}$ and $C_{rf}$) are used to realize the differential ports. Baluns ($L_m$ and $C_m$) are also used to provide the required anti-phase STM biases of the upper and lower SE circulators from three 120 deg phase-shifted modulation signals with an amplitude $V_m$ and frequency $f_m$. DC biasing is combined with these signals through the baluns' shunt inductance and a sufficiently large resistance $R_B$ connecting the two terminals of the baluns' balanced port. Furthermore, the parallel $LC$ tanks are realized using inductors $L_0$ and a pair of varactors (recall that each SE circulator in the voltage-mode topology is implemented using the bandstop/delta topology shown in Fig. 1(a)). The varactors are in common-cathode configuration and the common node is connected to an inductor $L_d$ so that they form together a bandpass resonance at $f_m$, thus allowing the modulation signal to pass through while, at the same time, prohibiting the RF signal from leaking out of the delta junction. Under the small-signal assumption, the circuit in Fig. 4(a) can be simplified as shown in Fig. 4(b) where the TX, RX and ANT ports along with the RF baluns are all replaced with differential voltage sources with a total impeance $Z_0$ (output impedance of the baluns). Also, the common-cathode varactors and the DC/modulation network are replaced with time-variant capacitors whose capacitance, assuming weak and linear modulation, is given by

$$C_n = \begin{cases} C_0 + \Delta C \cos(\omega_m t + \varphi_n), & n = 1, 2, 3 \\ C_0 - \Delta C \cos(\omega_m t + \varphi_n), & n = 4, 5, 6 \end{cases} \quad (6)$$

where $C_0$ is the static capacitance of the common-cathode varactors as set by the DC bias and $\Delta C$ is the effective capacitance variation which is proportional to the modulation voltage $V_m$. We also assume that the varactors' and the inductors' losses of each tank are combined into a dispersion-less parallel resistance $R_0 = Q_0 \omega_0 L_0$, where $Q_0$ is the unloaded quality factor of the tanks. Applying Kirchhoff's laws to the $n^{th}$ tank in Fig. 4(b) and writing the result in a matrix form, we get

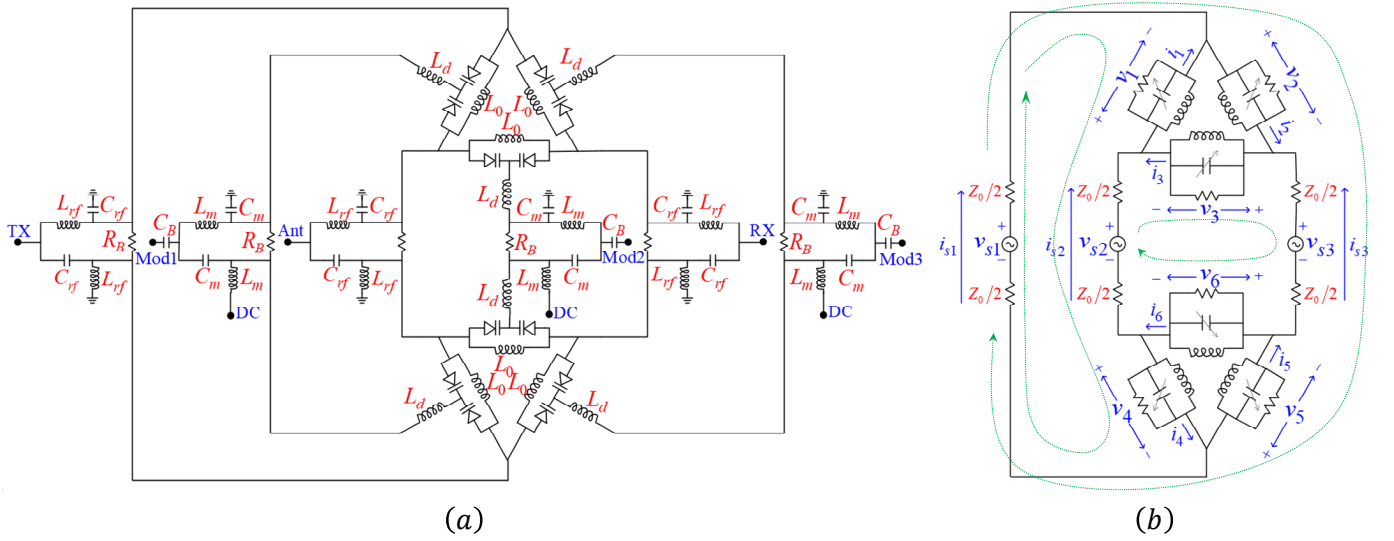

Fig. 4. Voltage-mode differential STM circulator: (a) Complete circuit implementation. (b) Small-signal model.

$$C_0 \overline{\overline{H}} \overline{v_d}'' + \left(\frac{1}{R_0}\overline{\overline{H}} - \frac{2}{Z_0}\overline{\overline{Q}}\right)\overline{v_d}' + \frac{1}{L_0}\overline{\overline{H}}\overline{v_d} + \Delta C \overline{\overline{H}} \overline{\overline{C_c}} \overline{v_c}'' \quad (7)$$
$$-\Delta C \omega_m \overline{\overline{H}} \overline{\overline{C_s}} \overline{v_c}' = -\frac{1}{Z_0}\overline{\overline{G}} \overline{v_s}',$$

$$C_0 \overline{v_c}'' + \frac{1}{R_0}\overline{v_c}' + \frac{1}{L_0}\overline{v_c} + \Delta C \overline{\overline{C_c}} \overline{v_d}'' - \Delta C \omega_m \overline{\overline{C_s}} \overline{v_d}' = 0, \quad (8)$$

where $' = \frac{d}{dt}$, $'' = \frac{d^2}{dt^2}$, $\overline{\overline{U}}$ is the unitary matrix, $\overline{\overline{H}}$, $\overline{\overline{Q}}$, $\overline{\overline{C_c}}$, and $\overline{\overline{C_s}}$ are matrix operators which are derived in Appendix A, $\overline{v_s} = \{v_{s1}, v_{s2}, v_{s3}\}$ is the differential source excitation vector, and $\overline{v_c}$ and $\overline{v_d}$ are the common and differential voltage vectors, respectively, given by

$$\overline{v_c} = \frac{\overline{v_u} + \overline{v_l}}{2}, \quad (9)$$

$$\overline{v_d} = \frac{\overline{v_u} - \overline{v_l}}{2}, \quad (10)$$

where $\overline{v_u} = \{v_1, v_2, v_3\}$ and $\overline{v_l} = \{v_4, v_5, v_6\}$ are the vectors of the tank voltages in the upper and lower SE STM circulators, respectively. Equations (7) and (8) can be further simplified if we express both $\overline{v_d}$ and $\overline{v_c}$ as a superposition of the junction's modes as expressed in (2). Applying this transformation and recognizing that the in-phase mode of each SE circulator is not excited (see Appendix A) yields

$$\frac{\Delta C}{2C_0}\begin{pmatrix}0 & e^{-j\omega_m t}\\ e^{+j\omega_m t} & 0\end{pmatrix}\begin{pmatrix}v_{c,+}''\\ v_{c,-}''\end{pmatrix} - j\frac{\Delta C}{2C_0}\omega_m\begin{pmatrix}0 & e^{-j\omega_m t}\\ -e^{+j\omega_m t} & 0\end{pmatrix}\begin{pmatrix}v_{c,+}'\\ v_{c,-}'\end{pmatrix}$$
$$+ \begin{pmatrix}v_{d,+}''\\ v_{d,-}''\end{pmatrix} + \frac{2R_0 + 3Z_0}{3R_0 Z_0 C_0}\begin{pmatrix}v_{d,+}'\\ v_{d,-}'\end{pmatrix} + \frac{1}{L_0 C_0}\begin{pmatrix}v_{d,+}\\ v_{d,-}\end{pmatrix} = \frac{-2/\sqrt{3}}{Z_0 C_0}\begin{pmatrix}e^{+j\pi/6}\\ e^{-j\pi/6}\end{pmatrix}v_{s1}', \quad (11)$$

$$\begin{pmatrix}v_{c,+}''\\ v_{c,-}''\end{pmatrix} + \frac{1}{R_0 C_0}\begin{pmatrix}v_{c,+}'\\ v_{c,-}'\end{pmatrix} + \frac{1}{L_0 C_0}\begin{pmatrix}v_{c,+}\\ v_{c,-}\end{pmatrix} + \frac{\Delta C}{2C_0}\begin{pmatrix}0 & e^{-j\omega_m t}\\ e^{+j\omega_m t} & 0\end{pmatrix}\begin{pmatrix}v_{d,+}''\\ v_{d,-}''\end{pmatrix}$$
$$-j\frac{\Delta C}{2C_0}\omega_m\begin{pmatrix}0 & e^{-j\omega_m t}\\ -e^{+j\omega_m t} & 0\end{pmatrix}\begin{pmatrix}v_{d,+}'\\ v_{d,-}'\end{pmatrix} = 0, \quad (12)$$

where $v_{d,\pm}$ and $v_{c,\pm}$ are the differential and common components of the counter-rotating modes, respectively. A circuit as in Fig. 4 with six first-order resonant tanks (three for each junction) would normally lead to a sixth-order system, but the absence of the in-phase modes reduces the order by two, as described by (11) and (12) which represent a fourth-order system of second-order linear differential equations. These equations can be solved by Fourier transform, which yields

$$\frac{V_{d,\pm}(\omega)}{V_{s1}(\omega)} = \frac{j\omega\, e^{\pm j\pi/6}}{3\sqrt{3}Z_0 C_0}\frac{\left[(\omega\pm\omega_m)^2 - \frac{1}{R_0 C_0}j(\omega\pm\omega_m) - \frac{1}{L_0 C_0}\right]}{D_\pm(\omega)}, \quad (13)$$

$$\frac{V_{c,\mp}(\omega\pm\omega_m)}{V_{s1}(\omega)} = \frac{-j(1\pm j/\sqrt{3})}{12}\frac{\frac{\Delta C}{Z_0 C_0^2}\omega^2(\omega\pm\omega_m)}{D_\pm(\omega)}, \quad (14)$$

where $V_{d,\pm}(\omega)$, $V_{c,\pm}(\omega)$, and $V_s(\omega)$ are the Fourier transforms of $v_{d,\pm}(t)$, $v_{c,\pm}(t)$, and $v_s(t)$, respectively, and

$$D_\pm(\omega) = -\left(\frac{\Delta C}{2C_0}\right)^2 \omega^2(\omega\pm\omega_m)^2 + \left[\omega^2 - \frac{2R_0 + 3Z_0}{3R_0 Z_0 C_0}j\omega - \frac{1}{L_0 C_0}\right]$$
$$\times\left[(\omega\pm\omega_m)^2 - \frac{1}{R_0 C_0}j(\omega\pm\omega_m) - \frac{1}{L_0 C_0}\right]. \quad (15)$$

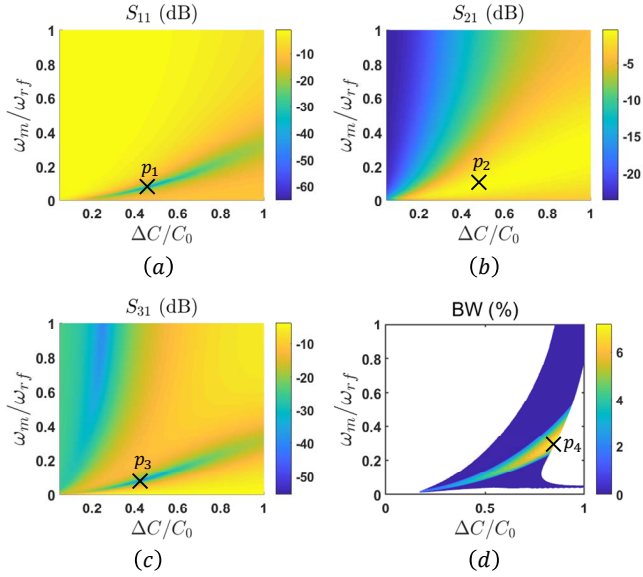

Fig. 5. *S*-parameters at 1 GHz versus modulation parameters for $Z_0 = 50\,\Omega$ and $Q_0 = 70$: (a) Return loss. (b) Insertion loss. (c) Isolation. (d) Fractional bandwidth.

Equations (13) and (14) show that the only frequency components existing in the circuit are the fundamental harmonics at $\omega$ and the first-order IM products at $\omega \pm \omega_m$ without any additional higher-order products, as incorrectly predicted by the approximate analysis of differential circuits based on half-circuit models in Sec. III.A. Also, the fundamental harmonics are exclusively excited as differential components, allowing them to flow to the ports while the first-order IM products are exclusively excited as common components, hence they are trapped inside the resonant junction and cannot carry power to the external ports. This result shows that, interestingly, the proposed circulator hides the intrinsic time-variant characteristics from the external ports, making it essentially a pseudo-LTI circuit. As mentioned earlier, this IM-free characteristic not only allows to achieve very low insertion loss, increases the effective power handling, and avoids interference with neighboring channels, but it also relaxes the required modulation parameters significantly as we show later in this paper. Finally, the *S*-parameters can be calculated from (13) as follows (see Appendix A)

$$S_{11}(\omega) = 2\left[-\frac{1}{6} - \frac{V_{d,+}(\omega) + V_{d,-}(\omega)}{V_{s1}(\omega)} + \frac{j}{\sqrt{3}}\frac{V_{d,+}(\omega) - V_{d,-}(\omega)}{V_{s1}(\omega)}\right], \quad (16)$$

$$S_{21}(\omega) = 2\left[\frac{1}{3} + \frac{V_{d,+}(\omega) + V_{d,-}(\omega)}{V_{s1}(\omega)} + \frac{j}{\sqrt{3}}\frac{V_{d,+}(\omega) - V_{d,-}(\omega)}{V_{s1}(\omega)}\right], \quad (17)$$

$$S_{31}(\omega) = 2\left[\frac{1}{3} - j\frac{2}{\sqrt{3}}\frac{V_{d,+}(\omega) - V_{d,-}(\omega)}{V_{s1}(\omega)}\right]. \quad (18)$$

Due to the circulator's threefold rotational symmetry, the rest of the *S*-parameters can be found by rotating the indices as $(1,2,3) \to (2,3,1) \to (3,1,2)$. With proper choice of the circuit elements and modulation parameters, $V_{d,\pm}$ can be designed to destructively interfere at one port and sum up at the other, as required to achieve infinite isolation. More specifically, port 3 can be isolated if $V_{d,\pm}/V_{s1} = \mp j/4\sqrt{3}$, which results in the following conditions on the modulation parameters when substituted in (13):

$$\frac{\left[(\omega_{rf} + \omega_m)^2 - \omega_0^2\right]\left[\omega_{rf}^2 + \frac{\sqrt{3}\omega_0}{Q_l}\omega_{rf} - \omega_0^2\right]}{\left[(\omega_{rf} - \omega_m)^2 - \omega_0^2\right]\left[\omega_{rf}^2 - \frac{\sqrt{3}\omega_0}{Q_l}\omega_{rf} - \omega_0^2\right]} = \frac{(\omega_{rf} + \omega_m)^2}{(\omega_{rf} - \omega_m)^2}, \quad (19)$$

$$\left(\frac{\Delta C}{2C_0}\right)^2 = \frac{\left[(\omega_{rf} \pm \omega_m)^2 - \omega_0^2\right]\left[\omega_{rf}^2 \pm \frac{\sqrt{3}\omega_0}{Q_l}\omega_{rf} - \omega_0^2\right]}{\omega_{rf}^2\left[\omega_{rf}^2 \pm \omega_m^2\right]}, \quad (20)$$

where $Q_l = Q_0 \| Q_r$ and $Q_r = \omega_0(3Z_0/2)C_0$ (recall that $Q_0 = \omega_0 R_0 C_0$ is the unloaded quality factor of the resonant tanks). For operation at a given frequency $\omega_{rf}$, (19) and (20) can be used to calculate the required modulation parameters $\omega_m$ and $\Delta C$ to achieve infinite isolation. Quite interestingly, the same conditions lead to unitary transmission at the third port, assuming a lossless circuit. Also, from power conservation, this leads to perfect matching at the input port. Therefore, the proposed differential circuits allow, in the lossless case, to realize an ideal circulator with *S*-matrix given by

$$\overline{\overline{S}} = \begin{bmatrix} 0 & 0 & 1 \\ 1 & 0 & 0 \\ 0 & 1 & 0 \end{bmatrix}, \quad (21)$$

while SE circulators do not have this advantage, since the required modulation parameters to optimize for return loss (RL), insertion loss (IL), and isolation (IX) are not the same [33]. In order to get further insight into the operation of the circulator, Fig. 5(a)-(c) show the *S*-parameters at the three ports for excitation from port 1 versus the modulation parameters $\omega_m$ and $\Delta C$ normalized to $\omega_0$ and $C_0$, respectively. These charts were generated using (16)-(18) assuming $Z_0 = 50\,\Omega$, $Q_0 = 70$, and an input frequency $f_{rf} = 1$ GHz. It can be seen that the required modulation parameters to optimize RL, IL, and IX as indicated by points $p_1$, $p_2$, and $p_3$, respectively, are nearly the same as mentioned earlier (the negligible misalignment is due to the finite $Q_0$). Moreover, Fig. 5(d) shows the circulator's BW, which is defined as the minimum frequency range to maintain IX more than 20 dB, RL less than 20 dB, and IL less than 3 dB. In general BW is a more relevant metric in practical systems than optimized *S*-parameters at a single-frequency. Optimizing the BW, however, requires different modulation parameters, as indicated by point $p_4$, than those at $p_1$, $p_2$, and $p_3$. Since the capacitance ratio at $p_4$ is impractical, we choose in this paper to operate at $\Delta C/C_0 = 0.5$ and $f_m/f_{rf} = 0.1$, which assumes the maximum realistic capacitance variation of commercial off-the-shelf varactors, hence results in the best possible BW, and at the same time is close to $p_2$ where IL becomes minimum. In such a case, Fig. 6(a) shows the theoretical *S*-parameters (based on (16)-(18)) where IL, RL, and IX at the center frequency are 0.6 dB, 27.5 dB, and 31.5 dB, respectively, and BW is 3.6% (36 MHz). We would like to stress that infinite IX at the center frequency is still possible if the circuit is designed at point $p_3$. However, in practice, it is desirable to minimize the dispersion of the *S*-parameters over the frequency range of operation, in order to avoid distortion of realistic signals with finite bandwidth. Uniform IX, in particular, is desirable as it simplifies the design of the following layers of self-interference cancellation in full-duplex systems. For the sake of comparison, Fig. 6(b) shows the *S*-parameters of a SE circulator designed using the same modulation frequency (capacitance ratio is optimized for lowest IL) where the results are clearly much worse. Specifically, IL, RL, and IX at 1 GHz all degrade to 2.31 dB, 21.8 dB, and 10 dB, respectively, and BW becomes undefined since the minimum level of 20 dB IX is not satisfied.

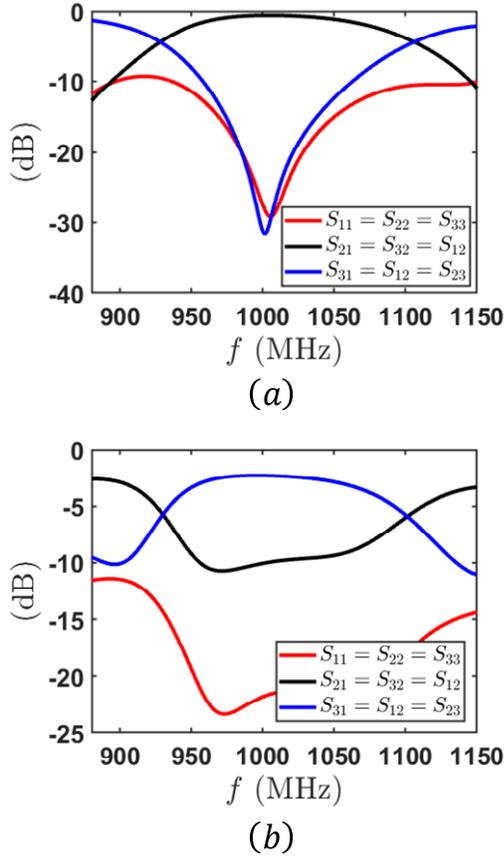

Fig. 6. Theoretical *S*-parameters for $f_m/f_{rf} = 0.1$ : (a) Differential. (b) Single-ended.

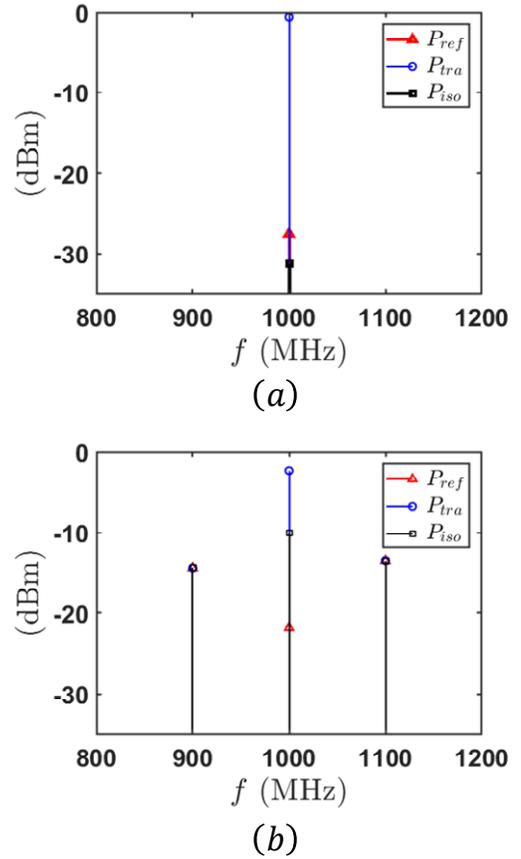

Fig. 7. Theoretical harmonic spectrum at all ports for an incident tone at $f_{rf} = 1$ GHz and $P_{in} = 0$ dBm: (a) Differential. (b) Single-ended.

As mentioned earlier, the drastic improvement in the differential architecture is due to the cancellation of IM products. Fig. 7(a) shows that the signal spectrum at all ports is indeed IM-free. Although the incident signal was assumed to be at $f_{rf} = 1$ GHz in Fig. 7(a), the same conclusion holds for any input frequency. In contrast, Fig. 7(b) shows that the IM products in the SE implementation are only –12 dBc and very close to the fundamental component (only ±100 MHz apart) which is the reason for the poor *S*-parameters in Fig. 6(b).

The analysis for the current-mode topology is analogous, and provided in Appendix B. Unlike their SE counterparts, the differential current-mode topology has several advantages compared to the differential voltage-mode topology. Specifically, since the baluns at the RF ports are eliminated, the overall insertion loss and noise figure can be improved, the form factor and complexity are reduced, and more immunity against random mismatches is maintained. Furthermore, the original disadvantages of the SE wye topology are eliminated, i.e., additional filters are not required in the modulation network to block the modulation signal, since the differential circuit exhibits virtual ground symmetry at the RF ports. Also, impedance transformers can be removed since the requirements on the modulation parameters and consequently the *Q*-factor of the circuit are all relaxed, thus allowing to design the current-mode circuit with 50 Ohm termination at a few GHz. However, the voltage-mode circulator still maintains an advantage in terms of linearity and power handling, since it relies on parallel LC tanks, which do not amplify the input voltage across the varactors, as the wye/current-mode topology, but rather they amplify the current, which does not contribute to the non-linearity of the varactors.

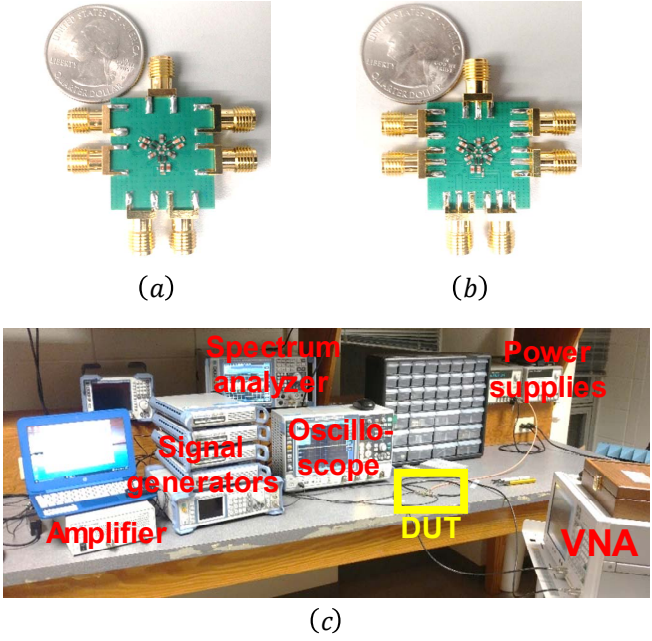

Fig. 8. Photograph of: (a) Top side of the fabricated prototype. (b) Bottom side of the fabricated prototype. (a) Experimental setup.

TABLE I
THEORETICAL DESIGN PARAMETERS

| Element | | Value |
|---|---|---|
| LC tanks | $D$ | ~10 pF @ VDC = 7.3 V |
| | $L_0$ | 4.3 nH |
| RF baluns | $L_{rf}$ | 11 nH |
| | $C_{rf}$ | 2.2 pF |
| Modulation baluns | $L_m$ | 56 nH |
| | $C_m$ | 43 pF |
| Biasing/Modulation network | $L_d$ | 150 nH |
| | $R_B$ | 100 KOhm |
| | $C_B$ | 1000 pF |

TABLE II
LIST OF THE USED EQUIPMENT

| Part | Model | Quantity |
|---|---|---|
| Power supply | Agilent E3631A | 1 |
| Vector network analyzer | Agilent E5071C | 1 |
| Spectrum analyzer | R&S FSVA40 | 1 |
| Oscilloscope | R&S RTO1044 | 1 |
| Signal generators | R&S SGS100A | 4 |
| | R&S SMB100A | 1 |
| Instrument amplifier | Minicircuits TVA-4W-422A+ | 1 |

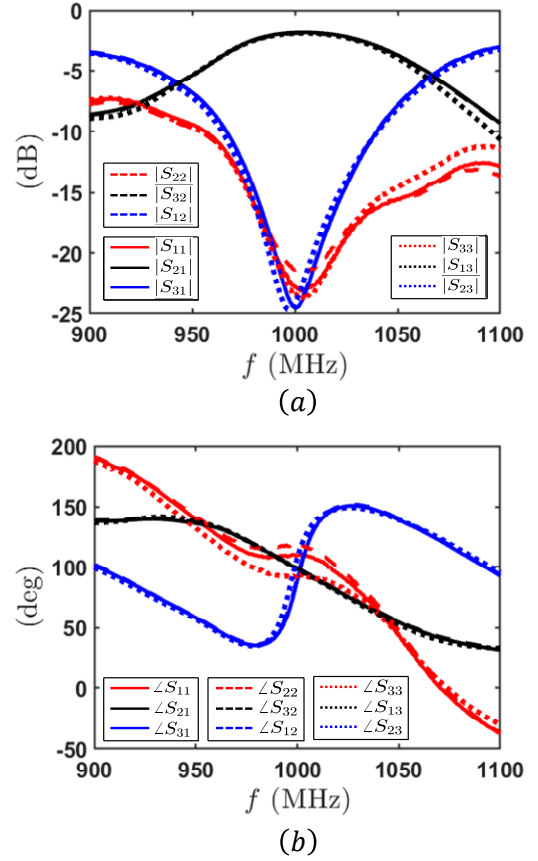

Fig. 9. Measured S-parameters: (a) Magnitude. (b) Phase.

## IV. RESULTS AND DISCUSSION

Guided by the theoretical analysis in Section III, a PCB prototype of the differential voltage-mode STM circulator (see Fig. 4(a) for the schematic) was designed at 1 GHz using off-the-shelf discrete components as listed in Table I. In order to account for all parasitics, the layout was simulated using ADS Momentum and the generated S-parameters were combined with commercially available spice models of all components to perform post-layout circuit/EM co-simulations. The design was then fabricated and the total area occupied by all components is 2×(13mm×11mm). Notice that both the top and bottom sides of the PCB are populated, each with one SE circulator to have a symmetric and more compact layout as shown in Figs. 8(a)-(b). Fig. 8(c) shows the measurement setup, while Table II provides a list of the used equipment to take the measurements.

### A. S-parameters

Fig. 9 shows the measured S-parameters in magnitude and phase from all ports for $V_{DC} = 7.3$ V, $V_m = 0.8$ V(rms), and $f_m = 100$ MHz. Notice that the modulation parameters, particularly the amplitude, are much lower than their SE counterparts [33], yet the differential circuit still results in much better performance. Specifically, the measured IL, RL, and IX at the circulator's center frequency of 1 GHz are 1.78 dB, 23 dB, and 24 dB, respectively, and the fractional BW is 2.3% (23 MHz). For the sake of comparison, the simulated

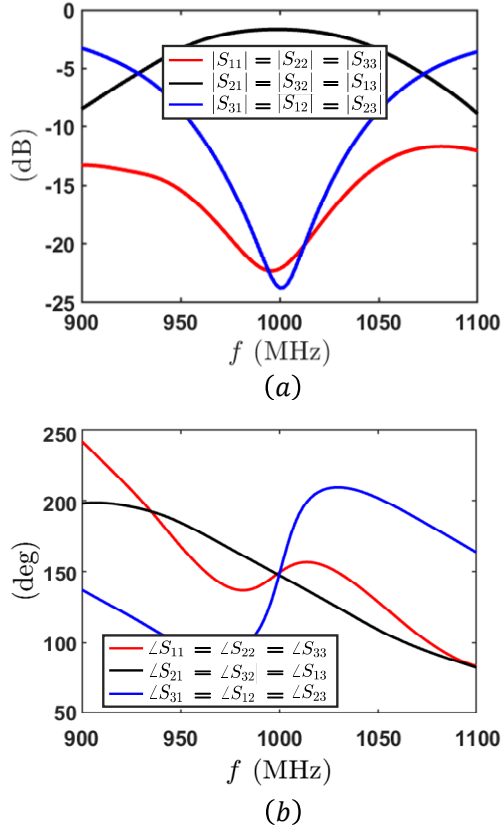

Fig. 10. Simulated *S*-parameters: (a) Magnitude. (b) Phase.

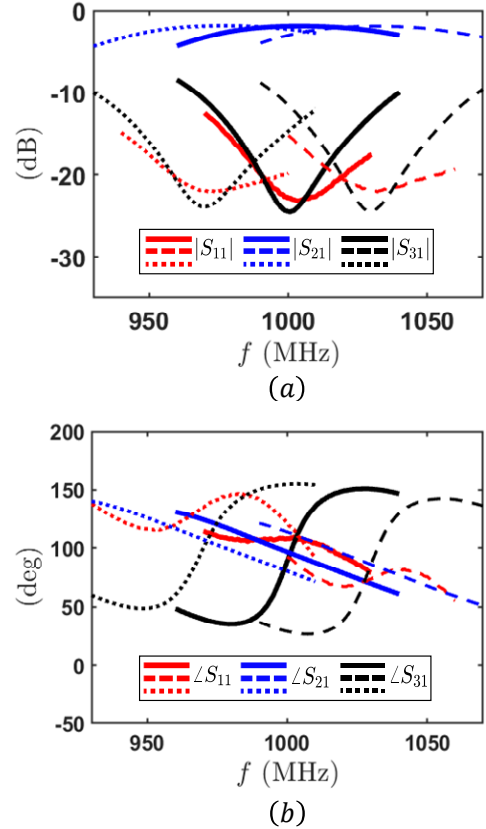

Fig. 11. Measured *S*-parameters at different channels by changing the DC bias and modulation voltage: (a) Magnitude. (b) Phase.

*S*-parameters are also shown in Fig. 10 where the simulated IL, RL, and IX at 1 GHz are 1.7 dB, 22 dB, and 24 dB, respectively, and the fractional BW is 2.3% (23 MHz). Clearly, simulated and measured *S*-parameters are in excellent agreement, yet they are different than the theoretical results in Fig. 6(a). The reason is that RF baluns were not included in the small-signal analysis of Section III.B for simplicity. These baluns contribute to insertion loss and, more importantly, result in a finite imbalance in the differential architecture due to random component mismatches. As one may expect, this imbalance is the reason for the slight asymmetry between different ports as shown in Fig. 9. The baluns' loss is estimated using circuit/EM co-simulations to be 0.5 dB each, which if de-embedded from the measured results, the actual IL of the differential circulator at 1 GHz becomes 0.78 dB which is in excellent with the theoretical results in Fig. 6(a). It is worth mentioning that this is the insertion loss that the circuit would exhibit if it were connected directly to a differential transceiver, in which case the baluns are omitted. To the best of our knowledge, this is the lowest IL of all LPTV magnetless circulators presented to-date, with available room for further improvement in a more optimized design or by sacrificing the low modulation parameters and increasing their values. Interestingly, the current-mode topology (see Fig. 15 and Appendix B), does not even require such baluns, therefore the total IL in this case can be expected to be below 1 dB. The circulator can also be tuned for operation at different channels as shown in Fig. 11 by simply controlling the DC bias of the varactors, and adjusting the modulation voltage accordingly in order to account for the different slope of the *CV* characteristics at the new quiescent point. The maximum tunability range, while maintaining the same specs on *S*-parameters, was measured to be 60 MHz (6% of the band center frequency at 1 GHz).

B. *Harmonic Response*

Fig. 12 shows the measured and simulated harmonic spectrum at both the transmitted ($P_{tra}$) and isolated ($P_{iso}$) ports for an incident signal at 1 GHz and 0 dBm. Despite that the measured IM products are as small as –29 dBc for a modulation frequency of only 10%, they are still finite, mainly due to the non-linear *CV* characteristics of the varactors (recall that linear time-variation produces no IM products in the proposed differential topologies, as shown in Sec. III). Random mismatches in the components, which are inevitable in practice, also play a role as they lead to finite imbalance not only in the RF baluns but more generally between the constituent SE circuits in any differential architecture. However, the impact of this imbalance is insignificant as can be deduced from the fact that the measured IM products are in excellent agreement with simulations where this issue is neglected. Compared to the SE architecture in [33], the IM products are at least 17 dB smaller (=1/50) even though $f_m$ is reduced from 190 MHz to only 100 MHz. We would like to stress that a SE implementation at 100 MHz would have resulted in even larger IM products and the overall performance, particularly isolation and insertion loss, would have become much worse.

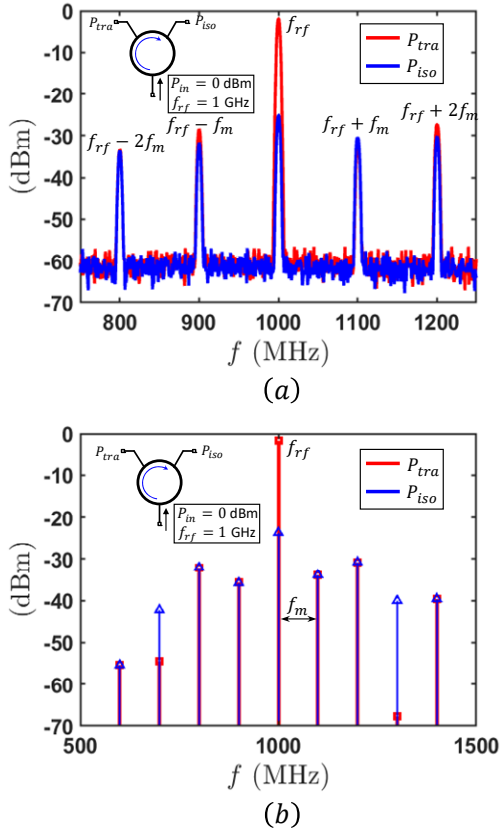

Fig. 12. Harmonic spectrum at transmitted and isolated ports for a single tone input at $f_{rf} = 1$ GHz and $P_{in} = 0$ dBm: (a) Measured. (b) Simulated.

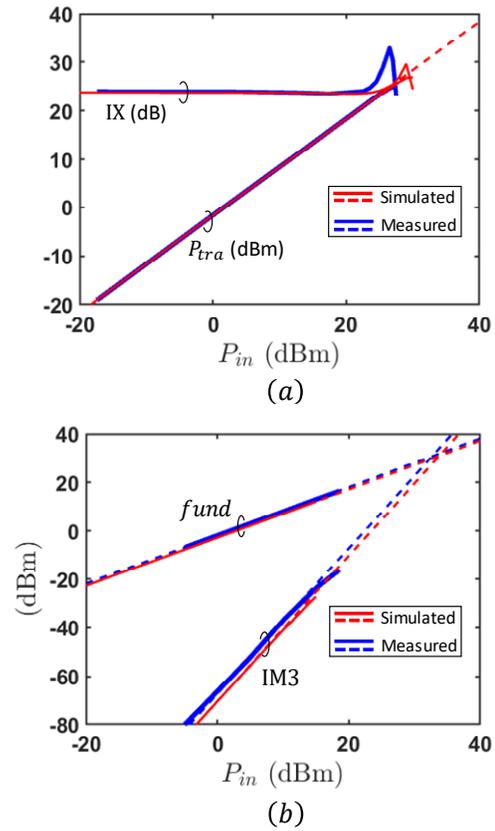

Fig. 13. (a) Measured and simulated P1dB and IX compression. (b) Measured and simulated IIP3.

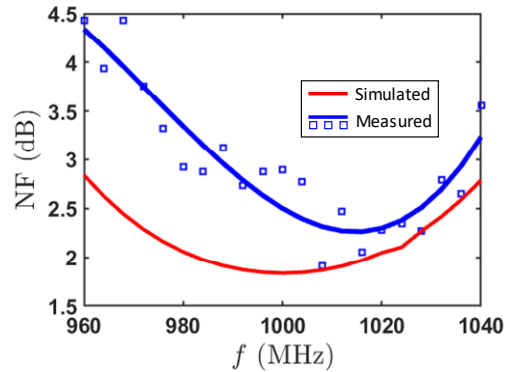

Fig. 14. Measured and simulated NF.

### C. Power Handling and Non-Linearities

Non-linearity in STM circulators is exclusively due to the varactors which ultimately leads to compression of both IL and IX. In [33], the maximum power (Pmax) that a circulator can handle was defined as the input power that guarantees an IL compression less than 1 dB (P1dB) and at the same time maintains IX more than 20 dB (IX20dB). Fig. 13(a) shows the measured and simulated results for both the transmitted power and isolation versus the input power at 1 GHz. The measured P1dB and IX20dB both occur at +28 dBm, thus resulting in Pmax of the same value, which is in good agreement with the simulated value of +30 dBm. We also notice a peaking in IX before Pmax since the capacitance variation $\Delta C$ at such high power is effectively reduced by the higher-order terms of the varactors' non-linear $CV$ characteristics, hence the operation point in the design charts shown in Fig. 5 is shifted down closer to $p_3$ where IX is maximum. Fig. 13(b) also shows the fundamental and third-order harmonic of the transmitted power for two in-band tones at 1 GHz with 1 MHz separation. The measured input-referred third-order intercept point (IIP3) is found to be +31 dBm and the simulated value is +33.8 dBm. The power handling of the presented circuit is slightly less than [33] solely because the varactors used in this paper have a lower breakdown voltage. If the same varactor were used in both SE and differential implementations, Pmax in the differential architecture would actually be 3 dB larger, since input power is halved between the upper and lower SE circulators through the baluns. Despite this artifact, the results of Fig. 13 are still larger than the reported values of any other magnetless circulator presented to-date.

### D. Noise Figure

In general, different mechanisms contribute to the NF of STM circulators [33]. This includes incoming noise from the RF ports which remains the same as in the SE architecture, assuming the typical 50 Ohm termination in both cases.

TABLE III
Summary of the Measured Results in Comparison to Previous Works.

| Metric | This work | [33] | [21] | [28] |
|---|---|---|---|---|
| Technology | SMT/PCB | SMT/PCB | CMOS/65 nm | SMA |
| RF center frequency (MHz) | 1000 | 1000 | 750 | 100 |
| Mod-to-RF frequency ratio (%) | 10 | 19 | 100 | 6 |
| DC bias (Volt) | 7.3 | 19.6 | N/A | N/A |
| Mod. amplitude (Vpp) | 2.3 | 10.8 | 1.2 | 8 |
| 20 dB IX BW (%) | 2.3 | 2.4 | 4.3 (†) | 200 |
| IL[1] (dB) | <2 (*) | <3.4 | <2 | 5 ~ 10 |
| RL[1] (dB) | >23 | >9 | N/A | >20 |
| P1dB[2] (dBm) | +28 | +29 | N/A (††) | N/A |
| IIP3[2] (dBm) | +31 | +33.7 | +27.5 (TX/ANT) +8.7 (ANT/RX) | N/A |
| NF[2] (dB) | 2.5 | 4.5 | 4 | N.A |
| First-order IM products (dBc) | –29 | –11.3 | N/A | N/A |
| Size (mm$^2$) | 2×(13×11) (**) | 13×11 | 5×5 | N/A |

[1] Over the BW.  *Baluns contribute 1 dB.  † Requires external impedance tuning.
[2] At center frequency.  **13×11 mm$^2$ on each side of the PCB.  †† Limited to +10 dBm by the breakdown voltage of the used 65nm technology.

However, noise generated by the circuit itself is doubled, very similarly to conventional differential circuits. In fact, the voltage-mode topology increases it even further because of additional loss in the RF baluns. Notice that the current-mode topology (Fig. 2(b) and Appendix B) does not have this problem. Amplitude and phase noise in the modulation sources also add to the total NF, yet if the anti-phase STM bias of the constituent SE circulators is generated from the same sources, as in this paper, then this noise becomes strongly correlated at the terminals of the differential ports and, therefore, cancels out. Noise folding from the IM frequencies also adds to the NF, but since the proposed differential circuits reduce these products, then their contribution becomes negligible. The reduction of the last two noise mechanisms, i.e., from the modulation sources and due to folding, is much stronger than the increased thermal noise generated by the circuit. Therefore, the overall NF improves. In fact, since the differential circuits are pseudo-LTI passive systems, the NF should be exactly equal to the IL [37]. Fig. 14 shows the measured and simulated NF, where the simulated value at 1 GHz is 1.8 dB, which is indeed nearly equal to the simulated IL, while the measured NF at 1 GHz is 2.5 dB which is about 0.7 dB larger than the measured IL due to imperfect cancellation of the IM products as discussed earlier. To the best of our knowledge, this is the smallest measured NF for magnetless circulators proposed to-date.

## V. CONCLUSION

We presented here a pseudo-LTI architecture for magnetless circulators, based on combining two bandstop/delta or bandpass/wye junctions with anti-phase STM bias in either a voltage- or a current-mode topology, respectively. We developed a rigorous analytical model for the proposed circuits and explained a detailed design procedure to achieve given specifications on insertion loss, return loss, isolation and bandwidth. Our theory shows that all IM products are indeed cancelled, despite the fact that the order of the system is increased because of the increased number of LC tanks. Based on this theory, we designed a PCB prototype for the voltage-mode topology and measured its performance including scattering parameters, harmonic response, power handling, and noise figure, as summarized in Table III in comparison with previous works. Several of these metrics surpass the results of all magnetless circulators presented to-date, with room for further improvement, thus making the differential STM circulator a significant step on the quest towards integrated full-duplex communication systems.

## APPENDICES

### A. Detailed Analysis of the Voltage-Mode Topology

Here, we present a detailed analysis for the differential voltage-mode topology and derive the equations given in section III.B. Applying Kirchhoff's laws to the $n$-th tank in Fig. 4(b) and following a similar analysis to [33], we get

$$\left(C_0\overline{\overline{U}} + \Delta C\overline{\overline{C_c}}\right)\overline{v_u}'' + \left(\frac{1}{R_0}\overline{\overline{U}} - \Delta C\omega_m \overline{\overline{C_s}}\right)\overline{v_u}' + \frac{1}{L_0}\overline{v_u} = \overline{i_u}', \quad (22)$$

$$\left(C_0\overline{\overline{U}} - \Delta C\overline{\overline{C_c}}\right)\overline{v_l}'' + \left(\frac{1}{R_0}\overline{\overline{U}} + \Delta C\omega_m \overline{\overline{C_s}}\right)\overline{v_l}' + \frac{1}{L_0}\overline{v_l} = -\overline{i_l}', \quad (23)$$

where $' = \frac{d}{dt}$, $\overline{i_u} = \{i_1, i_2, i_3\}$ and $\overline{i_l} = \{i_4, i_5, i_6\}$ are the vectors of the tank currents of the upper and lower SE STM circulators, respectively, $\overline{v_u} = \{v_1, v_2, v_3\}$ and $\overline{v_l} = \{v_4, v_5, v_6\}$ are the vectors of the tank voltages of the upper and lower SE STM circulators, respectively, $\overline{\overline{U}}$ is the unitary matrix, $\overline{\overline{C_c}}$ and $\overline{\overline{C_s}}$ are two matrices given by

$$\overline{\overline{C_c}} = \begin{bmatrix} \cos(\omega_m t) & 0 & 0 \\ 0 & \cos(\omega_m t + \alpha) & 0 \\ 0 & 0 & \cos(\omega_m t + 2\alpha) \end{bmatrix}, \quad (24)$$

$$\overline{\overline{C_s}} = -\frac{1}{\omega_m}\overline{\overline{C_c}}' = \begin{bmatrix} \sin(\omega_m t) & 0 & 0 \\ 0 & \sin(\omega_m t + \alpha) & 0 \\ 0 & 0 & \sin(\omega_m t + 2\alpha) \end{bmatrix},$$
(25)

and $\alpha = 2\pi/3$. The tank currents $\bar{i}_u$ and $\bar{i}_l$ can be related to the source currents $\bar{i}_s = \{i_{s1}, i_{s2}, i_{s3}\}$, which flow through the differential ports, using KCL as follows

$$\bar{i}_s = -\bar{\bar{G}}\bar{i}_u = \bar{\bar{G}}\bar{i}_l, \qquad (26)$$

where

$$\bar{\bar{G}} = \begin{bmatrix} +1 & -1 & 0 \\ 0 & +1 & -1 \\ -1 & 0 & +1 \end{bmatrix}. \qquad (27)$$

Next, the tank voltages $\bar{v}_u$ and $\bar{v}_l$ can be expressed as a superposition of differential and common components by rewriting (9) and (10) as follows

$$\bar{v}_u = \bar{v}_c + \bar{v}_d, \qquad (28)$$
$$\bar{v}_l = \bar{v}_c - \bar{v}_d. \qquad (29)$$

Substituting (26)-(28) into (22) and (23) yields

$$\bar{\bar{G}}\bar{v}_d'' + \frac{1}{R_0 C_0}\bar{\bar{G}}\bar{v}_d' + \frac{1}{L_0 C_0}\bar{\bar{G}}\bar{v}_d \\ + \frac{\Delta C}{C_0}\bar{\bar{G}}\bar{\bar{C}}_c\bar{v}_c'' - \omega_m \frac{\Delta C}{C_0}\bar{\bar{G}}\bar{\bar{C}}_s\bar{v}_c' = \frac{-1}{C_0}\bar{i}_s', \qquad (30)$$

$$C_0 \bar{v}_c'' + \frac{1}{R_0}\bar{v}_c' + \frac{1}{L_0}\bar{v}_c + \Delta C \bar{\bar{C}}_c \bar{v}_d'' - \Delta C \omega_m \bar{\bar{C}}_s \bar{v}_d' = 0. \qquad (31)$$

The source currents $\bar{i}_s$ can be related to the source voltages $\bar{v}_s$ by applying KVL along the loops shown in green dashed lines in Fig. 4(b), which results in

$$\bar{\bar{G}}\bar{i}_s = \frac{1}{Z_0}\bar{\bar{G}}\bar{v}_s - \frac{2}{Z_0}\bar{\bar{Q}}\bar{v}_d, \qquad (32)$$

where

$$\bar{\bar{Q}} = \bar{\bar{U}} - \bar{\bar{G}} = \begin{bmatrix} 0 & 1 & 0 \\ 0 & 0 & 1 \\ 1 & 0 & 0 \end{bmatrix}. \qquad (33)$$

Substituting (32) into (30) yields (7) and (8), where $\bar{\bar{H}}$ is given by

$$\bar{\bar{H}} = \bar{\bar{G}}\bar{\bar{G}} = \begin{bmatrix} 1 & -2 & 1 \\ 1 & 1 & -2 \\ -2 & 1 & 1 \end{bmatrix}. \qquad (34)$$

Equations (7) and (8) can be further simplified if we express $\bar{v}_d$ and $\bar{v}_c$ as a superposition of the in-phase, clockwise and counter-clockwise modes. This mode decomposition can be expressed through the following matrix transformation:

$$\tilde{v}_d = \bar{\bar{T}}^{-1}\bar{v}_d, \\ \tilde{v}_c = \bar{\bar{T}}^{-1}\bar{v}_c, \qquad (35)$$

where $\tilde{v}_d = \{v_{d,0}, v_{d,+}, v_{d,-}\}$ and $\tilde{v}_c = \{v_{c,0}, v_{c,+}, v_{c,-}\}$ are the vectors of the mode voltages for the differential and common components, respectively, and the operator $\bar{\bar{T}}$ is given by

$$\bar{\bar{T}} = \begin{bmatrix} 1 & 1 & 1 \\ 1 & e^{j\alpha} & e^{-j\alpha} \\ 1 & e^{j2\alpha} & e^{-j2\alpha} \end{bmatrix}. \qquad (36)$$

Applying this transformation to (7) and (8) yields

$$\tilde{\bar{H}}\tilde{v}_d'' + \left(\frac{1}{R_0 C_0}\tilde{\bar{H}} - \frac{2}{Z_0 C_0}\tilde{\bar{Q}}\right)\tilde{v}_d' + \frac{1}{L_0 C_0}\tilde{\bar{H}}\tilde{v}_d \\ + \frac{\Delta C}{C_0}\tilde{\bar{C}}_c\tilde{v}_c'' - \frac{\Delta C}{C_0}\omega_m\tilde{\bar{C}}_s\tilde{v}_c' = -\frac{1}{Z_0 C_0}\tilde{\bar{G}}\tilde{v}_c', \qquad (37)$$

$$\tilde{v}_c'' + \frac{1}{R_0 C_0}\tilde{v}_c' + \frac{1}{L_0 C_0}\tilde{v}_c + \frac{\Delta C}{C_0}\tilde{\bar{C}}_c\tilde{v}_d'' - \frac{\Delta C}{C_0}\omega_m\tilde{\bar{C}}_s\tilde{v}_d' = 0, \qquad (38)$$

where

$$\tilde{\bar{H}} = \bar{\bar{T}}^{-1}\bar{\bar{H}}\bar{\bar{T}} = 3\begin{bmatrix} 0 & 0 & 0 \\ 0 & e^{-j\pi/3} & 0 \\ 0 & 0 & e^{+j\pi/3} \end{bmatrix}, \qquad (39)$$

$$\tilde{\bar{Q}} = \bar{\bar{T}}^{-1}\bar{\bar{Q}}\bar{\bar{T}} = \begin{bmatrix} 1 & 0 & 0 \\ 0 & -e^{-j\pi/3} & 0 \\ 0 & 0 & -e^{+j\pi/3} \end{bmatrix}, \qquad (40)$$

$$\tilde{\bar{G}} = \bar{\bar{T}}^{-1}\bar{\bar{G}} = \frac{1}{\sqrt{3}}\begin{bmatrix} 0 & 0 & 0 \\ e^{-j\pi/6} & -e^{+j\pi/6} & j/2 \\ e^{+j\pi/6} & -e^{-j\pi/6} & -j/2 \end{bmatrix}, \qquad (41)$$

$$\tilde{\bar{C}}_c = \bar{\bar{T}}^{-1}\bar{\bar{H}}\bar{\bar{C}}_c\bar{\bar{T}} = \frac{3}{2}\begin{bmatrix} 0 & 0 & 0 \\ e^{j(\omega_m t - \pi/3)} & 0 & e^{-j(\omega_m t + \pi/3)} \\ e^{-j(\omega_m t - \pi/3)} & e^{j(\omega_m t + \pi/3)} & 0 \end{bmatrix}, \qquad (42)$$

$$\tilde{\bar{C}}_s = \bar{\bar{T}}^{-1}\bar{\bar{H}}\bar{\bar{C}}_s\bar{\bar{T}} = \frac{j3}{2}\begin{bmatrix} 0 & 0 & 0 \\ -e^{j(\omega_m t - \pi/3)} & 0 & e^{-j(\omega_m t + \pi/3)} \\ e^{-j(\omega_m t - \pi/3)} & -e^{j(\omega_m t + \pi/3)} & 0 \end{bmatrix}. \qquad (43)$$

Equations (37) and (38) can be reduced to (11) and (12) if we recognize that the in-phase mode of each SE circulator is not excited, i.e. $v_{d,0} = \dfrac{v_{u,0} - v_{l,0}}{2}$ and $v_{c,0} = \dfrac{v_{u,0} + v_{l,0}}{2}$ are both equal to zero, and assuming an excitation at only port 1, since a general $\overline{v}_s$ can be constructed using a linear superposition of individual port excitations. In order to find the S-parameters, (26) is also transformed into the basis of the rotating modes, resulting in

$$\begin{pmatrix} i_{s,+} \\ i_{s,-} \end{pmatrix} = \frac{1}{3Z_0} v_{s1} + \frac{1}{Z_0}\begin{pmatrix} 1 - j/\sqrt{3} & 0 \\ 0 & 1 + j/\sqrt{3} \end{pmatrix}\begin{pmatrix} v_{d,+} \\ v_{d,-} \end{pmatrix}. \quad (44)$$

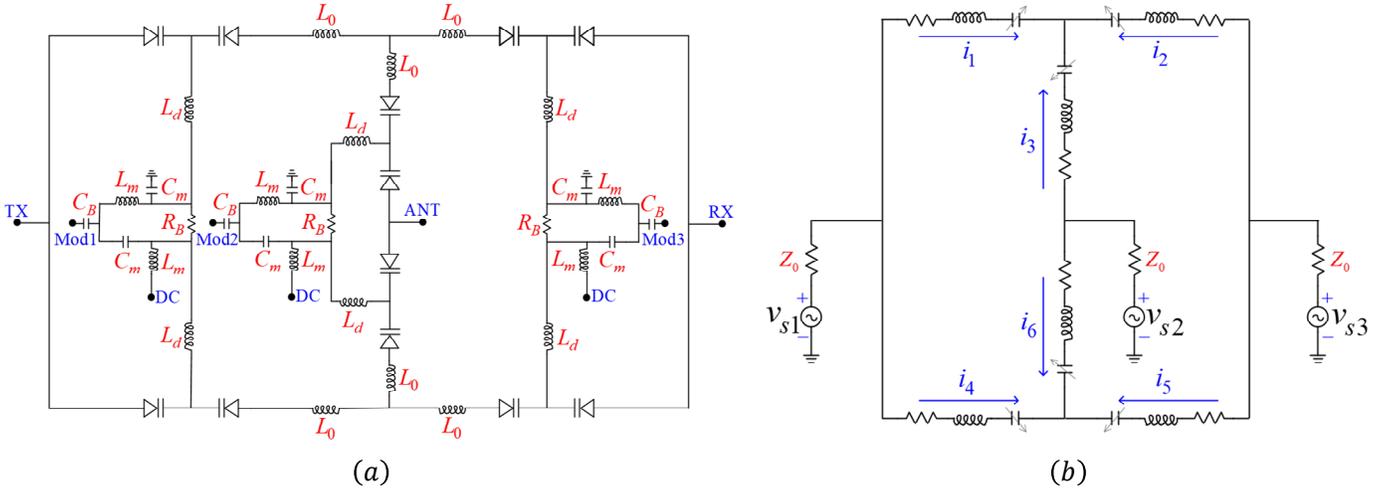

Fig. 15. Current-mode differential STM circulator: (a) Complete circuit implementation. (b) Small-signal model.

Fourier transforming (44) yields

$$I_{s,\pm}(\omega) = \frac{1}{3Z_0} V_{s1}(\omega) + \frac{1}{Z_0}\left(1 \mp j/\sqrt{3}\right)V_{d,\pm}(\omega), \quad (45)$$

where $V_{d,\pm}(\omega)$ are given by (13). Using the inverse transformation $\overline{I}_s = \overline{\overline{T}}\, \overline{I}_s$, the actual source currents $\overline{I}_s = \{I_{s1}, I_{s2}, I_{s3}\}$ are found as follows

$$I_{s1}(\omega) = I_{s,+}(\omega) + I_{s,-}(\omega), \quad (46)$$

$$I_{s2}(\omega_k) = e^{j\alpha} I_{s,+}(\omega) + e^{-j\alpha} I_{s,-}(\omega), \quad (47)$$

$$I_{s3}(\omega) = e^{-j\alpha} I_{s,+}(\omega) + e^{j\alpha} I_{s,-}(\omega). \quad (48)$$

By definition, the S-parameters are given by

$$S_{11}(\omega) = 1 - 2\frac{Z_0 I_{s1}(\omega)}{V_{s1}(\omega)},$$

$$S_{21}(\omega) = -2\frac{Z_0 I_{s2}(\omega)}{V_{s1}(\omega)}, \quad (49)$$

$$S_{31}(\omega) = -2\frac{Z_0 I_{s3}(\omega)}{V_{s1}(\omega)},$$

which can be simplified to (16)-(18).

### B. Detailed Analysis of the Current-Mode Topology

For completeness, we present the analysis of the dual differential current-mode topology (see Fig. 4(b)) in this appendix. The complete circuit implementation of this topology is shown in Fig. 15(a). Following similar steps as in Appendix A and applying Kirchhoff's laws to the n-th tank in Fig. 15(b), we get

$$L_0 \overline{i}_u'' + R_0 \overline{i}_u' + \frac{1}{3C_0}\overline{\overline{E}}\left(\sigma \overline{U} - \rho \overline{\overline{C}}_c\right)\overline{i}_u = \frac{1}{3}\overline{\overline{E}}\,\overline{v}_{in}', \quad (50)$$

$$L_0 \overline{i}_l'' + R_0 \overline{i}_l' + \frac{1}{3C_0}\overline{\overline{E}}\left(\sigma \overline{U} + \rho \overline{\overline{C}}_c\right)\overline{i}_l = \frac{1}{3}\overline{\overline{E}}\,\overline{v}_{in}', \quad (51)$$

where

$$\sigma = \frac{1}{\sqrt{1 - (\Delta C/C_0)^2}}, \quad (52)$$

$$\rho = \frac{2C_0}{\Delta C}\left(1 - \frac{1}{\sqrt{1 - (\Delta C/C_0)^2}}\right), \quad (53)$$

$$\overline{\overline{E}} = \begin{bmatrix} 2 & -1 & -1 \\ -1 & 2 & -1 \\ -1 & -1 & 2 \end{bmatrix}, \quad (54)$$

$\overline{i}_u = \{i_1, i_2, i_3\}$ and $\overline{i}_l = \{i_4, i_5, i_6\}$ are the vectors of the tank currents of the upper and lower SE STM circulators, respectively, $\overline{v}_{in} = \{v_1, v_2, v_3\}$ is the vector of the input voltages at the ports and is given by (24).

The tank currents $\overline{i}_u$ and $\overline{i}_l$ can, similarly, be expressed as a superposition of differential $\overline{i}_d$ and common $\overline{i}_c$ components as follows

$$\bar{i}_u = \bar{i}_c + \bar{i}_d, \quad (55)$$

$$\bar{i}_l = \bar{i}_c - \bar{i}_d. \quad (56)$$

Also, $\bar{v}_{in}$ can be related to the source voltages $\bar{v}_s$ using a simple KVL which gives

$$\bar{v}_{in} = \bar{v}_s - 2Z_0 \bar{i}_c, \quad (57)$$

Substituting from (55)-(57) into (50) and (51) yields

$$\bar{i}_d'' + \frac{R_0}{L_0}\bar{i}_d' + \frac{\sigma}{3L_0C_0}\bar{\bar{E}}\bar{i}_d - \frac{\rho}{3L_0C_0}\bar{\bar{E}}\bar{\bar{C}}_c\bar{i}_c = 0, \quad (58)$$

$$\bar{i}_c'' + \frac{3R_0 + 2Z_0}{3L_0}\bar{\bar{E}}\bar{i}_c' + \frac{\sigma}{3L_0C_0}\bar{\bar{E}}\bar{i}_c - \frac{\rho}{3L_0C_0}\bar{\bar{E}}\bar{\bar{C}}_c\bar{i}_d = \frac{1}{3L_0}\bar{\bar{E}}\bar{v}_s'. \quad (59)$$

Equations (58) and (59) can be transformed into the basis of the rotating modes $i_{d,\pm}$ and $i_{c,\pm}$ using the same matrix given by (36), which results in

$$\begin{pmatrix} i_{d,+}'' \\ i_{d,-}'' \end{pmatrix} + \frac{R_0}{L_0}\begin{pmatrix} i_{d,+}' \\ i_{d,-}' \end{pmatrix} + \frac{\sigma}{L_0C_0}\begin{pmatrix} i_{d,+} \\ i_{d,-} \end{pmatrix}$$
$$+ \frac{\rho}{2L_0C_0}\begin{pmatrix} 0 & e^{-j\omega_m t} \\ e^{+j\omega_m t} & 0 \end{pmatrix}\begin{pmatrix} i_{c,+} \\ i_{c,-} \end{pmatrix} = 0, \quad (60)$$

$$\begin{pmatrix} i_{c,+}'' \\ i_{c,-}'' \end{pmatrix} + \frac{R_0 + 2Z_0}{L_0}\begin{pmatrix} i_{c,+}' \\ i_{c,-}' \end{pmatrix} + \frac{\sigma}{L_0C_0}\begin{pmatrix} i_{c,+} \\ i_{c,-} \end{pmatrix}$$
$$+ \frac{\rho}{2L_0C_0}\begin{pmatrix} 0 & e^{-j\omega_m t} \\ e^{+j\omega_m t} & 0 \end{pmatrix}\begin{pmatrix} i_{d,+} \\ i_{d,-} \end{pmatrix} = \frac{1}{3L_0}v_{s1}'. \quad (61)$$

Notice that the in-phase modes $i_{d,0}$ and $i_{c,0}$ are equal to zero and we also assumed $\bar{v}_s = \{1,0,0\}$ for simplicity. Applying Fourier transform to (60) and (61) yields

$$\frac{I_{c,\pm}(\omega)}{V_{s1}(\omega)} = \frac{\frac{j\omega}{3L_0}\left[(\omega \pm \omega_m)^2 - \frac{R_0}{L_0}j(\omega \pm \omega_m) - \frac{\sigma}{L_0C_0}\right]}{D_\pm(\omega)}, \quad (62)$$

$$\frac{I_{d,\mp}(\omega \pm \omega_m)}{V_{s1}(\omega)} = \frac{\rho}{6L_0^2C_0}\frac{j\omega}{D_\pm(\omega)}, \quad (63)$$

where $I_{d,\pm}(\omega)$ and $I_{c,\pm}(\omega)$ are the Fourier transforms of $i_{d,\pm}(t)$ and $i_{c,\pm}(t)$, respectively, and $D_\pm(\omega)$ is given by

$$D_\pm(\omega) = \left(\frac{\rho}{2L_0C_0}\right)^2 + \left[-\omega^2 + \frac{R_0 + 2Z_0}{L_0}j\omega + \frac{\sigma}{L_0C_0}\right]$$
$$\times \left[(\omega \pm \omega_m)^2 - \frac{R_0}{L_0}j(\omega \pm \omega_m) - \frac{\sigma}{L_0C_0}\right]. \quad (64)$$

Applying KCL at the input terminals, the source currents $\bar{I}_s = \{I_{s1}, I_{s2}, I_{s3}\}$ can be found as follows:

$$\bar{I}_s = \bar{I}_u + \bar{I}_l = 2\bar{I}_c, \quad (65)$$

where $\bar{I}_c = \{I_{c1}, I_{c2}, I_{c3}\}$ is given by

$$I_{c1}(\omega) = \left[I_{c,+}(\omega) + I_{c,-}(\omega)\right], \quad (66)$$

$$I_{c2}(\omega) = \left[e^{j\alpha}I_{c,+}(\omega) + e^{-j\alpha}I_{c,-}(\omega)\right], \quad (67)$$

$$I_{c3}(\omega) = \left[e^{-j\alpha}I_{c,+}(\omega) + e^{j\alpha}I_{c,-}(\omega)\right]. \quad (68)$$

Finally, the *S*-parameters are calculated using (49) which yields

$$S_{11}(\omega) = 1 - \frac{4Z_0}{V_{s1}(\omega)}\left[I_{c,+}(\omega) + I_{c,-}(\omega)\right], \quad (69)$$

$$S_{21}(\omega) = \frac{2Z_0}{V_{s1}(\omega)}\left[I_{c,+}(\omega) + I_{c,-}(\omega) - j\sqrt{3}\left(I_{c,+}(\omega) - I_{c,-}(\omega)\right)\right], \quad (70)$$

$$S_{31}(\omega) = \frac{2Z_0}{V_{s1}(\omega)}\left[I_{c,+}(\omega) + I_{c,-}(\omega) + j\sqrt{3}\left(I_{c,+}(\omega) - I_{c,-}(\omega)\right)\right]. \quad (71)$$